\documentclass{jpsj-suppl}
\usepackage{txfonts} 
\renewcommand{\vec}[1]{\mbox{\boldmath $#1$}}

\title{
Beyond-mean-field approach to low-lying spectra of $\Lambda$ hypernuclei}

\author{Kouichi \textsc{Hagino}$^{1,2}$, 
Hua \textsc{Mei}$^{1,3}$, Jiangming \textsc{Yao}$^{1,3,4}$, and 
Toshio \textsc{Motoba}$^{5,6}$}

\inst{
$^{1}$Department of Physics, Tohoku University, Sendai 980-8578, Japan\\
$^2$
Research Center for Electron Photon Science, Tohoku University, 
1-2-1 Mikamine, Sendai 982-0826, Japan \\
$^3$School of Physical Science and Technology,
Southwest University, Chongqing 400715, China \\
$^4$
Department of Physics and Astronomy, University of North Carolina,
Chapel Hill, North Carolina 27516-3255, USA \\
$^5$Laboratory of Physics, Osaka Electro-Communication University,
Neyagawa 572-8530, Japan \\
$^6$
Yukawa Institute for Theoretical Physics, Kyoto University,
Kyoto 606-8502, Japan }

\recdate{}

\abst{
Taking the hypernucleus $^{13}_{~\Lambda}$C as an example, we illustrate 
the miscroscopic particle-rotor model for low-lying spectra of 
hypernuclei. This approach is based on the beyond-mean-field method, 
with the particle number and angular momentum projections. 
The quantum fluctuation of the mean-field is also taken into account 
for the core nucleus using the generator coordinate method. 
We show that the impurity effect of $\Lambda$ hyperon, such as 
a change in $B(E2)$, is well described 
with this model. Our calculation indicates that the most important 
impurity effect in $sd$-shell hypernuclei is a change in a deformation 
parameter rather than in a nuclear size.}

\kword{hypernuclei, low-lying spectrum, impurity effect, 
beyond-mean-field approach}

\begin{document}
\maketitle

\section{Introduction}

The development in $\Lambda$-hypernuclear spectroscopy has enabled
one to explore several
aspects of hypernuclear structure \cite{Hashimoto06}.
The measured energy spectra and electric multipole transition
strengths in low-lying states have in fact provided rich information
on the $\Lambda$-nucleon interaction in nuclear medium as well as on the
impurity effect of $\Lambda$ particle. 
Many theoretical methods have been developed to investigate the
spectroscopy  of hypernuclei, such as the 
cluster model \cite{Motoba83,Hiyama99}, 
the shell model \cite{Millener}, the ab-initio method \cite{abinitio},
the antisymmetrized molecular dynamics (AMD) \cite{Isaka11}, and
self-consistent mean-field models \cite{Zhou07,Win08,Win11,Lu11}. 
Among them, the self-consistent mean-field approach is the only
method which
can be globally applied from light to heavy hypernuclei.

Even though the self-consistent mean-field approach provides an 
intuitive view of nuclear deformation, 
it is a drawback of this method that 
it does not yield a spectrum in the laboratory frame, 
since the approach itself is formulated in the body-fixed
frame. This can actually 
be overcome by going beyond the mean-field approximation,
in particular, by carrying out the angular momentum projection.
One can also take into account the quantum fluctuation of the mean-field
wave function by superposing many Slater determinants with the
generator coordinate method (GCM). When the pairing correlation is
important, the particle number projection can also be implemented.
Such scheme has been referred to as a beyond-mean-field approach,
and has rapidly been developed in the nuclear structure physics
for the past decade \cite{Bender03,Yao14}.

In this contribution, we present 
a new method for low-lying states of hypernuclei
based on the beyond-mean-field approach \cite{Meihua1,Meihua2}.
In this method, 
we first apply the 
beyond-mean-field approach to a core nucleus. 
Low-lying states of hypernuclei are then constructed 
by coupling 
a $\Lambda$ particle to the core nucleus states. 
We thus call this approach the microscopic particle-rotor model, in 
which the rotor part is described with the microscopic beyond-mean-field 
method. See Refs. \cite{Meihua3,Weixia15,Zhou15,Yao11} 
for other different types of 
application of the beyond-mean-field approach to hypernuclei, 
which are complementary to the present microscopic particle-rotor model. 
We shall apply the microscopic particle-rotor model to the 
$^{13}_{~\Lambda}$C hypernucleus and discuss the impurity effect in this 
hypernucleus. 

\section{Microscopic particle-rotor model}

We consider a hypernucleus, which consists of a $\Lambda$ particle 
and an even-even core nucleus. 
In the microscopic particle-rotor model, 
the wave function for the whole $\Lambda$ hypernucleus 
with the angular momentum $J$ and its $z$-component $M$ 
is given as
\begin{equation}
 \label{wavefunction}
\Psi_{JM}(\vec{r},\{\vec{r}_N\})
 =\sum_{n,j,\ell, I}  R_{j\ell nI}(r) 
[{\cal Y}_{j\ell}(\hat{\vec{r}})\otimes
\Phi_{nI}(\{\vec{r}_N\})]^{(JM)},
\end{equation}
where 
$\vec{r}$ and $\vec{r}_N$ are the coordinates of the 
$\Lambda$ hyperon and the
nucleons, respectively. 
In this equation, 
${\cal Y}_{j\ell}(\hat{\vec{r}})$ is the spin-angular wave function 
for the $\Lambda$ hyperon, while 
$\vert\Phi_{nI}\rangle$ is the wave functions for the low-lying states
of the nuclear core. 
The latter is constructed from the mean-field wave 
functions as, 
\begin{equation}
\label{GCM}
 \vert \Phi_{nI M_I}\rangle
 =\int d\beta\,  f_{nI} (\beta)
\hat P^{I}_{M_IK} \hat P^N\hat P^Z\vert \varphi(\beta)\rangle, 
\end{equation}
where $\beta$ is the quadrupole deformation parameter and 
$|\varphi(\beta)\rangle$ is the mean-field wave function at $\beta$ 
obtained with the constrained mean-field approximation. 
Here we have assumed that the core nucleus has axial symmetric shape. 
$\hat P^{I}_{M_IK}$, $\hat P^N$, and $\hat P^Z$ are the projections 
operators for the angular momentum, the neutron number, and the 
proton number, respectively. 
The weight function  $ f_{nI}(\beta)$  in Eq. (\ref{GCM}) is
determined by the variational principle, that is, by solving the 
Hill-Wheeler equation. 

We assume that the total Hamiltonian for this system is given by, 
\begin{equation}
\hat H = \hat T_\Lambda + \sum^{A_c}_{i=1} v_{N\Lambda}(\vec{r},\vec{r}_{N_i}) 
+\hat{H}_{\rm c}, 
\label{eq:H}
\end{equation}
where $A_c$ is the mass number of the core nucleus.
Here, the first term 
is the kinetic energy of $\Lambda$ hyperon and the second term 
denotes a nucleon-hyperon interaction. 
The last term, $\hat{H}_{\rm c}$, is the Hamiltonian for the core 
nucleus, which is solved with the beyond-mean-field approach. 
With the Hamiltonian, Eq. (\ref{eq:H}), one can derive the 
coupled-channels equations for the radial wave functions, 
$R_{j\ell nI}(r)$, in which the coupling potentials are given 
in terms of the transition densities \cite{Meihua1,Meihua2}. 
The solutions of the coupled-channels equations 
provide the spectrum of the hypernucleus as well as the 
transition probabilities among the low-lying states. 

\section{Application to the $^{13}_{~\Lambda}$C hypernucleus}

We now apply the microscopic particle-rotor model to 
the $^{13}_{~\Lambda}$C hypernucleus, even though the model can be 
applied also to even heavier hypernuclei, such as $^{155}_{~~\Lambda}$Sm. 
To this end, we use 
the relativistic point coupling model with the PC-F1 
parameter set \cite{PC-F1} for the effective nucleon-nucleon 
interaction. For the nucleon-hyperon interaction, we use a 
simple relativistic zero-range interaction 
with a repulsive vector-type and an attractive scalar-type terms 
\cite{Meihua1,Meihua2}, 
\begin{equation}
v_{N\Lambda}(\vec{r},\vec{r}_{N})=\alpha_V^{N\Lambda}
\delta(\vec{r}-\vec{r}_N)
+\alpha_S^{N\Lambda}\gamma_\Lambda^0\delta(\vec{r}-\vec{r}_N) \gamma_N^0,
\end{equation}
where $\gamma^0$ is a Dirac matrix. 
The parameters 
$\alpha_V^{N\Lambda}$ and $\alpha_S^{N\Lambda}$ 
are determined so as to reproduce the empirical 
$\Lambda$ binding energy of 
$^{13}_{~\Lambda}$C. 
The coupled-channels equations for the radial wave functions, 
$R_{j\ell nI}(r)$, are solved by 
expanding 
$R_{j\ell nI}(r)$ 
on a spherical harmonic oscillator basis 
with 18 major shells. 
To this end, 
we include the 0$^+_1$, 2$^+_1$, 4$^+_1$, 
0$^+_2$, 2$^+_2$, and 4$^+_2$ states in the core nucleus, $^{12}$C. 

\begin{figure}[t]
\begin{center}
\includegraphics[scale=0.7,clip]{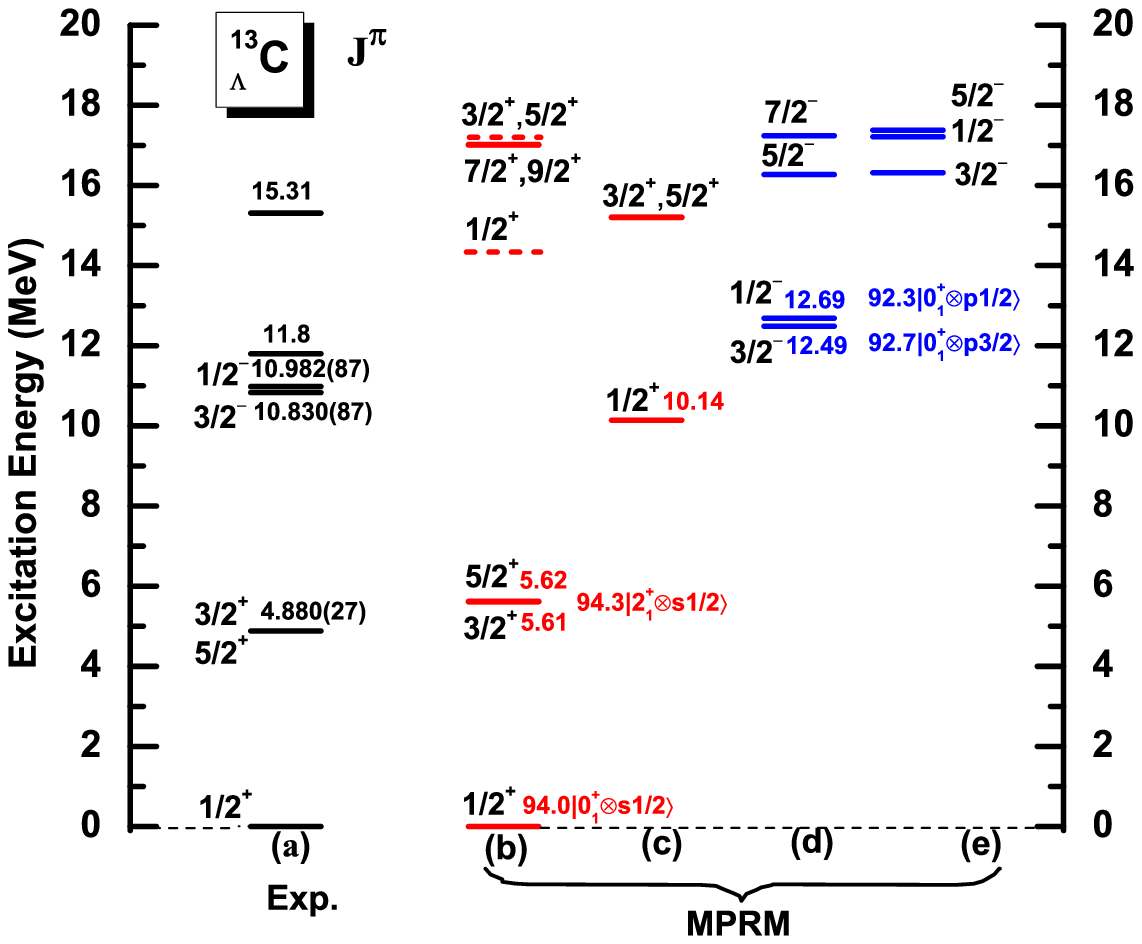}
\includegraphics[scale=0.6,clip]{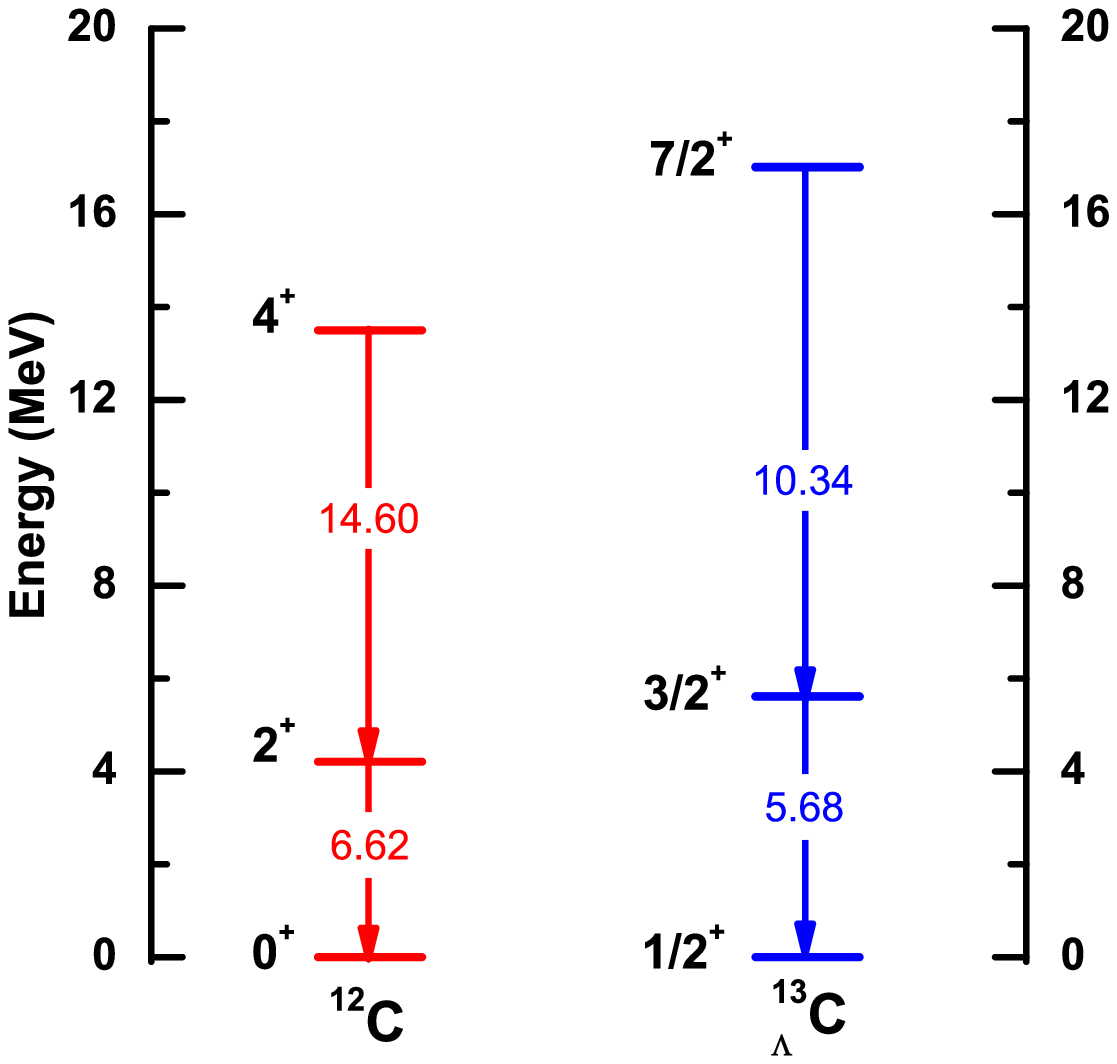}
\caption{The left panel: the spectrum of the $^{13}_{~\Lambda}$C hypernucleus 
obtained with the microscopic particle-rotor model. 
The experimental data are taken from Ref. \cite{Hashimoto06}. 
The right panel: a comparison of calculated $E2$ transition 
strengths, $B(E2)$, for $^{12}$C and $^{13}_{~\Lambda}$C, given in 
units of $e^2$ fm$^4$. 
}
\end{center}
\end{figure}

The left panel of Fig. 1 shows the low-lying spectrum of 
$^{13}_{~\Lambda}$C so obtained. 
One can see that the low-lying spectrum for $^{13}_{~\Lambda}$C is 
well reproduced,
although the excitation energies are slightly overestimated.
This calculation indicates the ground state rotational band of 
$^{13}_{~\Lambda}$C shown in the column (b), that is, 
the ground state 1/2$^+$ 
and the two doublets of $(5/2^+, 3/2^+)$ and $(9/2^+, 7/2^+)$, 
mainly consist of the configuration of $\Lambda s_{1/2}$ 
coupled to the ground rotational band (that is, 0$^+_1$, 2$^+_1$, and 
4$^+_1$) of the core nucleus, $^{12}$C. 
The doublet states are almost degenerate in energy, 
the energy difference being only 10 keV for each of the two doublets. 
The levels in the column (c) correspond to the configuration 
of $\Lambda s_{1/2}$ coupled to the second rotational 
band ($n=2$) in $^{12}$C. 
These states share similar features as those in the 
ground state band shown in the 
column (b).

In the negative-parity states shown in the column (d), 
the dominant configuration in the wave functions is that 
with the $\Lambda$ particle in the $p$ orbitals coupled to the ground 
state rotational band of the core nucleus. 
That is, 
the first $3/2^-$ and $1/2^-$ states consist mainly of 
$0_1^+\otimes\Lambda_{p_{3/2}}$ and $0_1^+\otimes\Lambda_{p_{1/2}}$, respectively, 
as is indicated in the figure. 
The energy splitting between these states is as small
as 199 keV, which 
reflects mainly
the spin-orbit splitting of $\Lambda$ hyperon 
in the $p_{3/2}$ and $p_{1/2}$ states.
The obtained splitting is in a good agreement with the empirical value, 
152$\pm$54$\pm$36 keV \cite{Ajimura01}. 

In contrast to the first $1/2^-$ and $3/2^-$ states, 
the second $1/2^-$ and $3/2^-$ states 
in the column (e) show 
a large configuration mixing. 
That is, the fraction of the 
$0_1^+\otimes\Lambda_{p_{1/2}}$ and 
$2_1^+\otimes\Lambda_{p_{3/2}}$ configurations 
in the 
wave function for the 1/2$^-_2$ state
is 0.60 and 0.38, respectively, 
while the 
fraction of the 
$2_1^+\otimes\Lambda_{p_{3/2}}$ and 
$2_1^+\otimes\Lambda_{p_{1/2}}$ configurations 
is 0.54 and 0.45, respectively, in the 
wave function for the 3/2$^-_2$ state \cite{Meihua2}. 
This large admixture of the configurations is 
due to the fact that 
there are two states whose unperturbed energy in the single-channel
calculations 
is close to one another.
A similar admixture occurs in other hypernuclei as well, 
such as $^9_\Lambda$Be and 
$^{21}_{~\Lambda}$Ne \cite{Meihua1,Meihua2}, which however show this feature 
already in the first $1/2^-$ and $3/2^-$ states. The difference between 
$^{13}_{\Lambda}$C and a pair of ($^9_\Lambda$Be, $^{21}_{~\Lambda}$Ne) originates 
mainly from the sign of the quadrupole deformation of the core nucleus, 
that is, an oblate deformation for $^{12}$C and a prolate deformation 
for $^8$Be and $^{20}$Ne \cite{Meihua2}. 

The right panel of Fig. 1 shows 
a comparison of the calculated $E2$ transition strengths for low-lying
positive parity states of $^{13}_{~\Lambda}$C with those of the core nucleus, 
$^{12}$C. 
In general, these transition strengths cannot be compared directly due to 
different angular momentum factors. However, 
for the transition from 
the 3/2$^+$ and the 5/2$^+$ states to the 1/2$^+$ state, 
such factor becomes trivial, and 
the transition strength can be directly interpreted as the that for the 
core nucleus from the 2$^+$ to the 0$^+$ states \cite{Meihua2}. 
Our calculation indicates that 
the $E2$ transition strength for
$2^+_1\rightarrow 0^+_1$ in $^{12}$C is significantly reduced, 
by a factor of $\sim14\%$, due to the addition of a $\Lambda$ particle.
The main cause of the reduction in the $B(E2)$ value is the reduction in 
nuclear deformation. According to our calculation, the proton radius $r_p$ 
is reduced from 2.44 fm to 2.39 fm by adding a $\Lambda$ particle 
to the $^{12}$C nucleus, which leads to about 7.9\% reduction in $r_p^4$. 
In contrast, the deformation parameter $\beta$ is altered from $-0.29$ 
to $-0.23$, leading to 37.1\% reduction in $\beta^2$. 
This clearly indicates that the change in deformation is the 
most important impurity effect in $sd$-shell hypernuclei. A similar 
conclusion has been reached also in Ref. \cite{Yao11}. 

\section{Summary}
 We have presented the microscopic particle-rotor model 
for the low-lying states of single-$\Lambda$ hypernuclei.
In this formalism, the wave functions for hypernuclei
are constructed by coupling the $\Lambda$ hyperon to 
the low-lying states of the core nucleus.
Applying this method to $^{13}_{~\Lambda}$C,
we have well reproduced the experimental energy spectrum
of this hypernucleus.
We have also found that the deformation is reduced 
by adding a $\Lambda$ particle in the positive-parity states, 
leading to a reduction in 
the $B(E2)$ value from the first $2^+$ to the ground states 
in the core nucleus. 

In this paper,
for simplicity, 
we have assumed the axial deformation for the core nucleus.  
An obvious extension of our method is to 
take into account
the triaxial deformation of the core nucleus.
One interesting application for this is
$^{25}_{~\Lambda}$Mg, for which
the triaxial degree of freedom has been shown to be important
in the core nucleus $^{24}$Mg. 

\section*{Acknowledgment}
This work was supported 
by JSPS KAKENHI Grant Number 2640263 and the NSFC under
Grant Nos. 11575148 and 11305134.

\end{document}